\documentclass[12pt]{article}
\usepackage{amsmath,amssymb,amsthm}
\usepackage{comment,url,verbatim}
\usepackage[usenames]{color}
\usepackage{hyperref}

\theoremstyle{definition}

\theoremstyle{remark}

\newtheorem*{remark*}{Remark}

\newcommand{\A}{\smallskip\noindent\texttt A:\  }
\newcommand{\Q}{\smallskip\noindent\texttt Q:\  }

\begin{document}

\title{Inverse Privacy}
\author{Yuri Gurevich\quad Efim Hudis\quad Jeannette M. Wing\\[0.5em] Microsoft Research, Redmond, WA, USA}
\date{}
\maketitle

\begin{abstract}
An item of your personal information is inversely private if some party has access to it but you do not. We analyze the provenance of inversely private information and its rise to dominance over other kinds of personal information. In a nutshell, the inverse privacy problem is unjustified inaccessibility to you of your inversely private information. We argue that the inverse privacy problem has a market-based solution.
\end{abstract}

\section{Introduction}
\label{sec:intro}

Call an item of your personal information inversely private if some party has access to it but you do not. The provenance of your inversely private information can be totally legitimate.
Your interactions with various institutions---employers, municipalities, financial institutions, health providers, police, toll roads operators, grocery chains, etc.---create numerous items of personal information, e.g., shopping receipts and refilled prescriptions. Due to progress in technology, institutions have become much better than you in recording data. As a result, shared data decays into inversely private. More inversely private information is produced when institutions analyze your private data.

Your inversely private information, whether collected or derived, allows institutions to serve you better. But access to that information---especially if it were presented to you in a convenient form---would do you much good. It would allow you to correct possible errors in the data, to have a better idea of your health status and your credit rating, and to identify ways to improve your productivity and quality of life.

In some cases, the inaccessibility of your inversely private information can be justified by the necessity to protect the privacy of other people and to protect the legitimate interests of institutions.
We argue that there are numerous scenarios where the chances to hurt other parties are negligible.
The inaccessibility of your inversely private information in such safe scenarios is the inverse privacy problem. A good solution of the problem should not only provide you accessibility to your inversely private information but should also make that access convenient.

We analyze the root causes of the inverse privacy problem and discuss a market-based solution for it. We concentrate here on the big picture, leaving many finer points for later analysis.

In \S\ref{sec:infoset}, we define your personal information. Our definition is relatively narrow but appropriate for the problem at hand. In \S\ref{sec:classify}, we give an access-based classification of your personal information. In \S\ref{sec:provenance}, we discuss the provenance of personal information. In \S\ref{sec:rise}, we discuss the rise of inversely private information to dominance over the other kinds of personal information. In \S\ref{sec:entitled}, we discuss the inaccessibility of your inversely private information, and we posit the Inverse Privacy Entitlement Principle. In \S\ref{sec:ipp}, we formulate and discuss the inverse privacy problem. Finally, in \S\ref{sec:predict}, we discuss a market-based solution of the inverse privacy problem and related issues.

Some explanations are more natural in a dialog, and so the reader will find below some discussions between Quisani, ostensibly a former student of the first author, and the authors, speaking one at at time.

\section{Personal infoset}
\label{sec:infoset}

For brevity, items of information are called \emph{infons} \cite{G198}.

An infon is \emph{tangible} if it has a material embodiment, e.g., written down on a piece of paper or recorded in some database. The same infon (as an abstract item of information) may have distinct material embodiments. Herein we restrict attention to infons that are tangible.

We are interested in scenarios where a person interacts with an institution, e.g. a shop, a medical office, a government agency. We say that an infon $x$ is \emph{personal} to an individual $P$ if (a)~$x$ is related to an interaction between $P$ and an institution and (b)~$x$ identifies $P$. A typical example of such an infon is a receipt for a credit-card purchase by a customer in a shop.

Define the \emph{personal infoset} of an individual $P$ to be the collection of all infons personal to $P$. Note that the infoset evolves over time. It acquires new infons. It may also lose some infons. But, because of the tangibility restriction, at any given moment, the infoset is finite.

\begin{quote}

\Q Give me an example of an intangible infon.

\A A fleeting impression that you have of someone who just walked by you.

\Q What about information announced but not recorded at a meeting? One can argue that the collective memory of the participants is a kind of embodiment.

\A Such a case of unrecorded information becomes less and less common. People write notes, write emails, tweet, send messages, use their smartphones to make videos, etc. Companies tend to tape their meetings. Numerous sensors, such as cameras and microphones, are commonplace and growing in pervasiveness, even in conference rooms. But yes, there are border cases as far as tangibility is concerned. At this stage of our analysis, we abstract them away.

\Q In the shopping receipt example, the receipt may also mention the salesclerk that helped the customer.

\A The clerk represents the shop on the receipt.

\Q But suppose that something went wrong with that particular purchase, the customer complained that the salesclerk misled her, and the shop investigates. In the new context, the person of interest is the salesclerk. The same infon turns out to be personal to more than one individual.

\A This is a good point. The same infon may be personal to more than one individual but we are interested primarily in contexts where the infon in question is personal to one individual.

\end{quote}

\section{Classification}
\label{sec:classify}

The personal infoset of an individual $P$ naturally splits into four buckets.

\begin{enumerate}
\item The \emph{directly private bucket} comprises the infons that $P$ has access to but nobody else does.

\item The \emph{inversely private bucket} comprises the infons that some party has access to but $P$ does not.

\item The \emph{partially private bucket} comprises the infons that $P$ has access to and a limited number of other parties do as well.

\item The \emph{public bucket} comprises the infons that are public information.
\end{enumerate}

\begin{quote}
  \Q Why do you call the second bucket ``inversely private''?

  \A The Merriam-Webster dictionary defines ``inverse'' as ``opposite in order, nature, or effect.'' The description of bucket~2 is the opposite of that of bucket~1.

  \Q As far as I can see, you discuss just two dimensions of privacy: whom a given infon is personal to, and who has access to the infon. The world is more complex, and there are other dimensions to privacy. Consider for example the pictures in the directly private bucket of my infoset that are personal to me only. Some of the pictures are clearly more private than others; there are degrees of privacy.

  \A Indeed we restrict attention to the two dimensions. But this restricted view is informative, and it allows us to carry on our analysis. Recall that we concentrate here on the big picture leaving many finer points for later analysis.

  \Q Concerning the public bucket of my infoset, how can public information be personal? Personal and public are the opposites.

  \A Recall the two privacy dimensions that we do take into account. There is no contradiction between being personal in one dimension and public in another.

\end{quote}

\section{Provenance}
\label{sec:provenance}

With time, the personal infoset of an individual acquires new infons. She may create new infons on her own, e.g., by making a selfie, by writing down some observation, or by writing down some conclusions that she inferred from information available to her.

But the infoset acquires many more new infons due to the interactions of the individual with other parties. The other parties could be people, e.g., relatives, neighbors, coworkers, clerks, waiters, and medical personnel. They could be institutions, e.g., employers, banks, internet providers, brick and mortar shops, online shops, and government agencies. The new infons could be factual records, gossip, rumors, or derived information.

The infoset may also lose some infons, especially if they have a unique embodiment. For example, the individual may destroy old letters or delete a selfie without sending it to anybody. Institutions also may lose or delete (embodiments of) infons, but in general, these days, institutions are much better then people in keeping records.

New items of a personal infoset do not necessarily stay in the bucket where they arose. Because of modern superiority of institutional bookkeeping, there is a flow of information from the partial privacy bucket to the inverse privacy bucket. In the next section we look into this dynamics.

\section{The rise of inverse privacy to dominance}
\label{sec:rise}

People have always interacted among themselves, and people have interacted with institutions for a very long time, certainly from the times that ancient governments started to collect taxes.
Until recently the capacity of a person to take and keep records was comparable to that of institutions.  Yes, the government kept tax records but, by and large, the people knew about their taxes as much as the government did.
Traditionally, the partial privacy bucket easily dominated the inverse privacy bucket.

Later on, governments, especially dictatorial governments, could marshal resources to collect information on people; a novelist illustrated this power the best \cite{Orwell}. The most radical change, however, is due to technology introduced in the last 20--30 years. The capacity of public and private institutions to take and keep records became vastly superior to that of a regular person. As a result, the large majority of items in the personal infoset is now generated as inversely or partially private. Often infons start as partially private but then quickly decay into inversely private because the institutions remember it all while the person often hardly remembers that the interaction took place.

For a regular citizen of an advanced society today, the volume of the inverse privacy bucket vastly exceeds that of the partial privacy bucket. Of course it may be simplistic to count bits or even items. A picture of a car has many bits but only so much useful information; even many pictures of the same car may have only so much useful information. It makes more sense to speak about the value of information rather than its volume.

Determining the value of personal information is a difficult problem, particularly because of a gap between what people are willing to pay for keeping an item of information directly private and what they are willing to accept for sharing that same item of information; see \cite{Acquisti} and its references. Nevertheless, we posit that typically the value of the inverse privacy bucket exceeds that of the partial privacy bucket and grows much faster.

Thus, in advanced societies today, the inverse privacy bucket of a typical personal infoset dominates the whole infoset. We see the dominance of inverse privacy as a problem. In this connection, it is important to understand legal, political, sociological, ethical, technological implications of the inverse privacy domination.

It is worth emphasizing that the main reason that we live now in world dominated by inverse privacy is not the invasion of privacy (the tremendous importance of that issue notwithstanding) but the gross disparity in the capability to take and keep records.

\section{The inverse privacy entitlement principle}
\label{sec:entitled}

Enterprises have legitimate reason to collect data about their customers; this allows them to serve their customers better. Medical institutions have legitimate reasons to collect data about their patients; this helps them diagnose and treat diseases. Governments have legitimate reason to collect data about their citizens; this helps them address societal problems.

As noted earlier, institutions are much better than individuals in collecting data. So, in the process of all the collection of data about customers, patients, and citizens, partially private data is quickly becoming inversely private. Aside from any surreptitious collection of personal information, this conversion of data from partially private to inversely private is critical to the provenance of inversely private information.

Access to your inversely private infons would allow you to correct possible errors in the data, to have a better idea of your health status and your credit rating, and so on and so forth.

From an ethical point of view, it is only fair to give you access to your personal infons. Already the 1973 HEW report \cite{hew} advocated that ``[t]here must be no personal-data record-keeping systems whose very existence is secret,'' and ``[t]here must be a way for an individual, to find out what information about him is in a record and how it is used.''
And the 1970 Fair Credit Reporting Act (FCRA) stipulated that, subject to various technical exceptions, ``[e]very consumer reporting agency shall, upon request, \dots clearly and accurately disclose to the consumer'' all information in the consumer's file, the sources of the information, etc. \cite[\S609]{FCRA}.

Concentrating on the big picture, we ignore technical exceptions here. But we cannot ignore that governments have legitimate security concerns, and businesses have legitimate competition concerns. The 2012 Federal Trade Commission (FTC) report on Protecting Consumer Privacy in an Era of Rapid Change'' is more nuanced: ``Companies should provide reasonable access to the consumer data they maintain; the extent of access should be proportionate to the sensitivity of the data and the nature of its use'' \cite[p.~64]{FTC2012}.
To this end, we posit

\medskip\noindent{\bf The Inverse Privacy Entitlement Principle}\quad
As a rule, individuals are entitled to access their personal infons.
There may be exceptions, but each such exception needs to be justified, and the burden of justification is on the proponents of the exception.

\medskip
One obvious exception is related to national security. The proponents of that exception, however, would have to justify it. In particular they would have to justify which parts of national security fall under the exception.

\section{The inverse privacy problem}
\label{sec:ipp}

We say that an institution \emph{shares back} your personal infons if it gives you access to them. This technical term will make the exposition easier.

Institutions may be reluctant to share back personal information, and they may have reasonable justifications: the privacy of other people needs to be protected, there are security concerns, there are competition concerns.
But there are numerous \emph{safe scenarios} where the chances are negligible that sharing back your personal infons would violate the privacy of another person or damage the legitimate interests of the information holding institution or any other institution.

The \emph{inverse privacy problem} is the inaccessibility to you of your personal information in such safe scenarios.

\begin{quote}
\Q Give me examples of safe scenarios.

\A Your favorite supermarket has plentiful data about your shopping there. Do you have that data?

\Q No, I don't.

\A But, in principle you could have. So how can sharing that data with you hurt anybody?
Similarly, many other businesses and government institutions have information about you that you could in principle have but in fact you do not. Some institutions share a part of your inversely private information with you but only a part. For example, Fitbit sends you weekly summaries but they have much more information about you.

\Q As you mentioned earlier, institutions have not only raw data about me but also derived information. I can imagine that some of that derived information may be sensitive.

\A Yes, there may be a part of your inversely private information that is too sensitive to be shared with you. Our position is, however, that the burden of proof is on the information holding institution.

\Q You use judicial terminology. But who is the judge here?

\A The ultimate judge is society.

\Q Let me raise another point. Enabling me to access my inversely private information makes it easier for intruders to find information about me.

\A This is true.  Any technology invented to allow inverse privacy information to be shared back has to be made secure.  Communication channels have to be secure, encryption has to be secure, etc. Note, however, that today hackers are in a much better position to find your inversely private information about you than you are. Sharing that information with you should improve the situation.
\end{quote}

\section{Going forward}
\label{sec:predict}

As we pointed out above, the inverse privacy problem is not simply the result of ill will of governments or businesses. It is primarily a side effect of technological progress. Technology influences the social norms of privacy \cite{Warren-Brandeis}. In this particular case, technology created the problem, and technology is instrumental in solving it.
In this section, we argue that the inverse privacy problem can be solved and will be solved. By default we restrict attention to safe scenarios of \S\ref{sec:ipp}.

\subsection{Social norms}
\label{sub:norms}

Individuals would greatly benefit from gaining access to their inversely private infons. They will have a much fuller picture of their health, their shopping history, places they visited, and so on. Besides, they would have an opportunity to correct possible errors in inversely private infons.

To what extent do people understand the great benefits of accessing their inversely private infons? We do not have data on the subject. Here's one indication \cite[\S4.2.2]{Lorry}.

\begin{quote}
  We asked participants to “[t]hink about the ability to view and edit the information that advertising companies know about you. How much do you agree or disagree with the following,” showing them six statements. 90\% of participants believed (agreed, strongly agreed) that they should be given the opportunity to view and edit their profiles. A large percentage wanted to be able to decide what advertising companies can collect about them (85\%) and saw benefits in being able to view (79\%) and edit profiles (81\%). The majority thought that the ability to edit their proﬁles would provide companies with more accurate data (70\%) and allow them to better serve the participants (64\%).
\end{quote}

As people realize the benefits in question more and more,  they will demand access to their inversely private infons louder and louder. Indeed, it is easy to underestimate the amount of information that businesses about a regular citizen. The story of Austrian privacy activist Max Schrems is instructive. ``In 2011, Schrems demanded that Facebook give him all the data the company had about him. This is a requirement of European Union (EU) law. Two years later, after a court battle, Facebook sent him a CD with a	1,200-page PDF'' \cite[Part~1, \S1]{Schneier}.

Social norms will evolve accordingly, toward a broad acceptance of the Inverse Privacy Entitlement Principle of \S\ref{sec:entitled}. Institutions should share back personal information as a matter of course. Furthermore, they should do so in a convenient way. Your personal infons should be available to you routinely and easily---just as the photos that you upload to a reputable cloud store. You do not have to file a legal request to obtain them.

The evolving social norms influence the law, and the law helps to shape social norms. For brevity, we restrict attention to the US law. We already quoted, in \S\ref{sec:entitled}, the 1970 Fair Credit Reporting Act, the 1973 HEW report, and the 2012 FTC report. See also the 2000 report of the FTC Advisory Committee on Online Access and Security \cite{FTC2000}, the 2003 Fair and Accurate Credit Transactions Act \cite{FACTA}, California's ``Shine the light'' law of 2003 \cite{CA_1798.83}, and the 2014 FTC report ``Data Brokers: A Call for Transparency and Accountability'' \cite{FTC2014}. Clearly the law favors transparency and facilitates your access to your inverse private infons.

\subsection{Market forces}
\label{sub:market}

The sticky point is whether companies will share back our personal information. This information is extremely valuable to them. It gives them competitive advantages, and so it may seem implausible that companies will share it back. We contend that companies will share back personal information because \emph{it will be in their business interests}.

Sharing back personal information can be competitively advantageous as well. Other things being equal, wouldn't you prefer to deal with a company that shares your personal infons with you?  We think so. Companies will compete on (a)~how much personal data, collected and derived, is shared back and (b)~how convenient that data is presented to customers.

The evolution toward sharing back personal information seems slow. This will change. Once some companies start sharing back personal data as part of their routine business, the competitive pressure will quickly force their competitors to join in. The competitors will have little choice.

There is money to be made in solving the inverse privacy problem. As sharing back personal information gains ground, the need will arise to mine large amounts of customers' personal data on their behalf. The benefits of owning and processing this data will grow, especially as the data involves financial and quality-of-life domains.
We anticipate the emergence of a new market for companies that compete in processing large sets of private data for the benefits of the data producers, i.e. consumers.

The miners of personal data will work on behalf of consumers and compete on how helpful they are to the customers, how trustworthy they are.
This emerging market will generate its own pressure on the personal data holders and potentially might find ways to benefit them as well. For example, if you shop at some retailer $R$ your personal data miner $M$ may show you a separate webpage devoted to $R$, suggest ways for you to save money as you shop there, and show you how $R$ intends to improve your shopping experience. The last part may even be written by $R$, but---working on your behalf---$M$ may also suggest to you better deals or shopping experiences elsewhere. The retailer $R$ will benefit if it can beat the competition.

\subsection{Better record keeping}
\label{sub:tools}

Finally, technology can enhance people's capacity to take and keep records. For example, your smartphone or wearable device may eventually become a trusted and universal recorder of many things you do.  Technology will help people maintain a personal diary effortlessly.

The project ``Small Data'' lead by Deborah Estrin at Cornell Tech \cite{Estrin2013} pioneers such an approach in the domain of health.
``Consider a new kind of cloud-based app that would create a picture of your health over time by continuously, securely, and privately analyzing the digital traces you generate as you work, shop, sleep, eat, exercise, and communicate'' \cite{Estrin2014}.

The ``small'' in ``Small Data'' reflects the fact that the personal health-related data of one individual isn't \emph{big data} \cite{Big_data}. In contrast to Estrin's work, we do not restrict attention to any particular data vertical. In our case, inversely private data of an individual may be on the biggish side; recall the story of Max Schrems above. In fact, ``Biggish'' is the name of a technical proposal, and this paper grew up out of Biggish ideas. We intend to address other Biggish ideas elsewhere.


\begin{thebibliography}{99}

\bibitem{Acquisti} Alessandro Acquisti, Leslie John, and George Loewenstein, ``What is Privacy Worth?'' in Privacy Papers for Policy Makers 2010,\\ \url{http://www.futureofprivacy.org/privacy-papers-2010/}

\bibitem{CA_1798.83} California S.B. 27, ``Shine the Light Law,'' Civil Code \S1798.83,\\ \url{http://goo.gl/zxuHUi}

\bibitem{Estrin2013} Deborah Estrin, ``What happens when each patient becomes their own "universe" of unique medical data?'' TedMed 2013,\\
\url{http://www.tedmed.com/talks/show?id=17762}

\bibitem{Estrin2014} Deborah Estrin, ``Small Data where $n=$ me,'' Communications of the ACM, Vol.\ 57 No.\ 4, Pages 32--34 (2014)

\bibitem{FACTA} Fair and Accurate Credit Transaction Act of 2003, Public Law 108--159, 108th Congress, p. 117,
    \url{http://goo.gl/PnpCkv}

\bibitem{FCRA} Fair Credit Reporting Act (FCRA), 1970. Title 15 U.S. Code, sec. 1681, \url{http://www.ftc.gov/os/statutes/fcra.htm}

\bibitem{FTC2000} Federal Trade Commission, ``Final Report of the FTC Advisory Committee on Online Access and Security,'' 2000\\
    \url{http://govinfo.library.unt.edu/acoas/papers/finalreport.htm}

\bibitem{FTC2012} Federal Trade Commission, ``Protecting Consumer Privace in an Era of Rapid Change,'' March 2012, \url{http://tinyurl.com/p6r3cy4}

\bibitem{FTC2014} Federal Trade Commission, ``Data Brokers: A Call for Transparency and Accountability,'' May 2014,
    \url{http://alturl.com/dwwy5}

\bibitem{G198} Yuri Gurevich and Itay Neeman, ``Infon Logic: the Propositional Case,'' \emph{ACM Transactions on Computation Logic} 12:2, article 9, January 2011

\bibitem{hew} U.S. Department of Health, Education, and Welfare (HEW), ``Records, Computers, and the Rights of Citizens,'' Report of the Secretary's Advisory Committee on Automated Personal Data Systems (the HEW report), July 1973,
    \url{https://epic.org/privacy/hew1973report/}.

\bibitem{Lorry} Pedro Giovanni Leon, Ashwini Rao, Florian Schaub,
Abigail Marsh, Lorrie Faith Cranor, and Norman Sadeh, ``Why People Are (Un)willing to Share Information with Online Advertisers,'' Carnegie Mellon University, Computer Science Technical Report CMU-ISR-15-106,\\
\url{http://goo.gl/RkqkhJ}

\bibitem{Nissenbaum} Helen Nissenbaum, ``Privacy in Context: Technology, Policy and the Integrity of Social Life,'' Stanford University Press, 2010.

\bibitem{Orwell} George Orwell, ``Nineteen Eighty-Four,'' Secker and Warburg, London, 1949.

\bibitem{Schneier} Bruce Schneier, ``Data and Goliath: The hidden buttles to collect your data and control your world,'' Noton and Company, 2015.

\bibitem{Strom} David Strom, ``Deborah Estrin wants to (literally) open source your life,'' ITworld, May 24, 2013, \url{http://goo.gl/Jsd9CT}

\bibitem{Warren-Brandeis} Samuel D. Warren and Louis D. Brandeis, ``The Right to Privacy,'' Harvard Law Review 4:5 (1890), 193--220.

\bibitem{Big_data} Wikipedia, ``Big data'' (8 August 2015), retrieved 9 August 2015,\\ \url{https://en.wikipedia.org/?oldid=675136834}


\end{thebibliography}
\end{document}